\documentclass[%
 preprint,
superscriptaddress,
 amsmath,amssymb,
 aip,jap,
]{revtex4-1}
\UseRawInputEncoding
\usepackage{graphicx}
\usepackage{dcolumn}
\usepackage{bm}
\usepackage{siunitx}
\usepackage{verbatim}
\usepackage{hyperref}

\usepackage{color,soul}
\usepackage{epstopdf}
\AppendGraphicsExtensions{.tif}

\begin{document}


\title{Effect of Buffer Termination on Intermixing and Conductivity in LaTiO$_3$/SrTiO$_3$ Heterostructures Integrated on Si(100) }
\author{Tongjie Chen}
 \affiliation{Department of Physics, North Carolina State University, Raleigh, North Carolina 27695, USA}

\author{Kamyar Ahmadi-Majlan}
 \affiliation{Department of Physics,University of Texas-Arlington, Arlington, TX, 76019 USA}

\author{Zheng Hui Lim}
 \affiliation{Department of Physics,University of Texas-Arlington, Arlington, TX, 76019 USA}

\author{Zhan Zhang}
 \affiliation{Advanced Photon Source, Argonne, IL, 76019 USA}

\author{Joseph H. Ngai}
 \affiliation{Department of Physics,University of Texas-Arlington, Arlington, TX, 76019 USA}
 
\author{Divine .P. Kumah}
\email{dpkumah@ncsu.edu}
\affiliation{Department of Physics, North Carolina State University, Raleigh, North Carolina 27695, USA}

\date{\today}

\begin{abstract}
The control of chemical exchange across heterointerfaces formed between ultra-thin functional transition-metal oxide layers provides an effective route to manipulate the electronic properties of these systems. We show that cationic exchange across the interface between the Mott insulator, LaTiO$_3$(LTO) grown epitaxially on SrTiO$_3$(STO)-buffered Silicon by molecular beam epitaxy depends strongly on the surface termination of the strained STO buffer. Using a combination of temperature-dependent transport and synchrotron X-ray crystal truncation rods and reciprocal space mapping, an enhanced conductivity in STO/LTO/SrO- terminated STO buffers compared to heterostructures with TiO$_2$-terminated STO buffers is correlated with La/Sr exchange and the formation of metallic La$_{1-x}$Sr$_x$TiO$_3$. La/Sr exchange effectively reduces the strain energy of the system due to the large lattice mismatch between the nominal oxide layers and the Si substrate. 
\end{abstract}

\maketitle

\section{Introduction}

Structural, electronic and chemical interactions at the interfaces between ultra-thin complex perovskite oxides can lead to exciting physical properties which are not found in the bulk analogue materials including interfacial magnetism, two-dimensional electron gases, metal-insulator transitions and superconductivity.\cite{hellman2017interface, chakhalian2012whither, ngai_correlated_2014} The ability to control these interactions at heterointerfaces using atomic layer-by-layer synthesis techniques such as molecular beam epitaxy allows for the unprecedented tailoring of electronic and magnetic ground states. Structural coupling including  strain, oxygen octahedral distortions and interfacial polar distortions can be induced by epitaxial constraints provided by appropriate substrate and buffer layers.\cite{ pentcheva2009avoiding, jia2009oxygen, zubko_interface_2011, rondinelli2012control} Additionally, thermodynamic and kinetic effects can lead to chemical interdiffusion across heterointerfaces leading to significant changes to composition away from nominal values which can significantly alter their physical properties.\cite{nakagawa2006some, qiao2013impacts} For example, intermixing between the nominally insulating LaTiO$_3$ (LTO) and SrTiO$_3$ (STO) layers can lead to the formation of a conducting La$_x$Sr$_{1-x}$TiO$_3$ (LSTO) interface in addition to the charge transfer mechanism proposed to alleviate the divergent electric field which arises due to the polar discontinuity at the LTO/STO interface.\cite{tokura_filling_1993, veit2019three} Interfacial intermixing at polar/non-polar interfaces has been proposed as a contributing mechanism to the high-mobility two dimensional electron gas formed at the interface between LaAlO$_3$ and SrTiO$_3$ \cite{willmott_structural_2007a, qiao2010thermodynamic, lee_hidden_2016, ohtomo_highmobility_2004, nakagawa2006some, chambers2011understanding, pentcheva2010electronic} and interfacial conductivity in LaCrO$_3$/SrTiO$_3$ superlattices.\cite{comes2017probing}

Of particular interest is understanding how the interfacial structure and electronic behavior of epitaxial oxides evolve when integrated on Si. Oxides exhibit a variety of properties that can potentially be exploited in device applications, provided integration onto Si is achieved.\cite{kumah_epitaxial_2020, hu2003interface, woicik_anomalous_2006, mi2008atomic, chambers2001band, kornblum_oxide_2015} The surface unit-cell of Si(100) has a lattice constant of 3.84 {\AA}, which in many cases will impart compressive strain on an epitaxial oxide. Intermixing across a heterointerface may be energetically favorable if it leads to a reduction of the strain energy of a system.\cite{sankara2014misfit, lipinski2000strain}  Developing ways to either promote or minimize intermixing, depending on its desirability, is crucial for controlling the functionality of epitaxial oxides on Si.

In this letter, we explore the effects of deposition sequence and chemical termination of a buffer layer on intermixing at polar/non-polar interfaces within STO/LTO/STO trilayers grown on Si(100). We find that deposition sequence and chemical termination have pronounced effects on intermixing and electronic behavior in heterostructures that are nominally identical in terms of composition. To determine the degree of chemical intermixing at the LTO/STO interface and the effect on the transport properties of the heterostructures, we investigate the atomic-scale structures of 1.5 uc STO cap/ 3 uc LTO/ N uc STO buffer/(001) Si samples grown by molecular beam epitaxy where the STO buffer thickness N is either 6 unit cells  or 4 unit cells. (1 unit cell (uc) =1 monolayer (ML) La(Sr)O + 1 ML TiO$_2$) We find that in addition to strain-driven chemical intermixing,  La-Sr exchange is more favorable if the LTO is grown by co-depositing LaO and TiO$_2$ on SrO-terminated STO buffers leading to higher conductivity compared to samples where the STO is TiO$_2$-terminated. The structural profiles are obtained by synchrotron-diffraction based crystal truncation rod measurements and high-resolution reciprocal space maps. 

Bulk LTO is a Mott insulator with a pseudo-cubic lattice constant of 3.96 \AA{}.\cite{cwik_crystal_2003} The Mott insulating state can be broken resulting in metallicity by epitaxial strain, over-oxidation or doping with divalent Sr.\cite{zhang2019mott, he2012metal, wong_metallicity_2010, ishida_origin_2008, hays_electronic_1999} Bulk STO is a band insulator with a pseudocubic lattice constant of 3.905 \AA{}. A high mobility two-dimensional electron gas (2DEG) forms at the LTO/STO interface exhibiting superconductivity.\cite{seo_optical_2007,biscaras_twodimensional_2010} Interest in integrating the unique properties of the oxide 2DEG system with semiconductor-based technologies has led to the fabrication of LTO/STO heterostructures on Si and Ge.\cite{jin_high_2014} STO grows epitaxially on (001)-oriented Si with the epitaxial relationship given by STO[110]/Si[100] and STO[001]//Si[001]. The c-axis of STO and Si are aligned in the out-of-plane direction with the perovskite STO lattice rotated in-plane by 45$^\circ$ with respect to the Si surface.\cite{mckee_crystalline_1998, reiner_crystalline_2010}, The in-plane lattice mismatch between STO and Si is given by $\epsilon = \frac{a_{\text{STO}}-a_{\text{Si surface}}}{a_{\text{Si}}}$ where the in-plane lattice spacing of STO, $a_{\text{STO}}$=3.905 \AA{} and the in-plane lattice spacing of the Si (001) surface $a_{\text{Si surface}}$=3.84 \AA{}. Given these values, the calculated STO/Si lattice mismatch is 1.66 \%. Due to the large lattice mismatch and the step-structure of the Si surface, strain relaxation of the STO lattice is known to occur within a critical thickness, $t_{\text{critical}}$= 5-10 unit cells.\cite{kumah_atomic_2010, warusawithana_ferroelectric_2009, chen_interfacial_2018, woicik_anomalous_2006, segal_morphology_2010}.

By controlling the thickness of the STO buffer, $t_{STO_b}$, the strain state of the LTO layers can be effectively tuned. For LTO films grown on thin STO buffer layers with $t_{STO}$ less than$t_{critical}$, the lattice mismatch between the LTO films and the STO buffer coherently strained to Si is 2.5 \% and thus, it becomes energetically favorable to reduce the strain energy by relaxing the strain through the formation of dislocations, and/or by chemical intermixing with the STO layers to form La$_x$Sr$_{1-x}$TiO$_3$ which has a smaller bulk lattice constant than LTO.\cite{hays_electronic_1999}

\section{Experiment}
The nominal 1.5 uc STO cap/ 3 uc LTO/ N uc STO buffer/(001) Si heterostructures were grown by molecular beam epitaxy. 2'' diameter Si (100)-oriented wafers (Virginia Semiconductor) were loaded into the MBE chamber and cleaned by exposing to activated oxygen generated by a radio frequency source to remove residual organics from the surface. To desorb the native oxide layer which formed at the surface of the Si substrate, 2 monolayers of Sr were deposited at a substrate temperature of 550 $^\circ$C, and the sample was heated to 870 $^\circ$C to remove the native layer of SiO$_x$ through the formation and desorption of SrO. 

\begin{figure}[ht]
\includegraphics[width=0.75\textwidth]{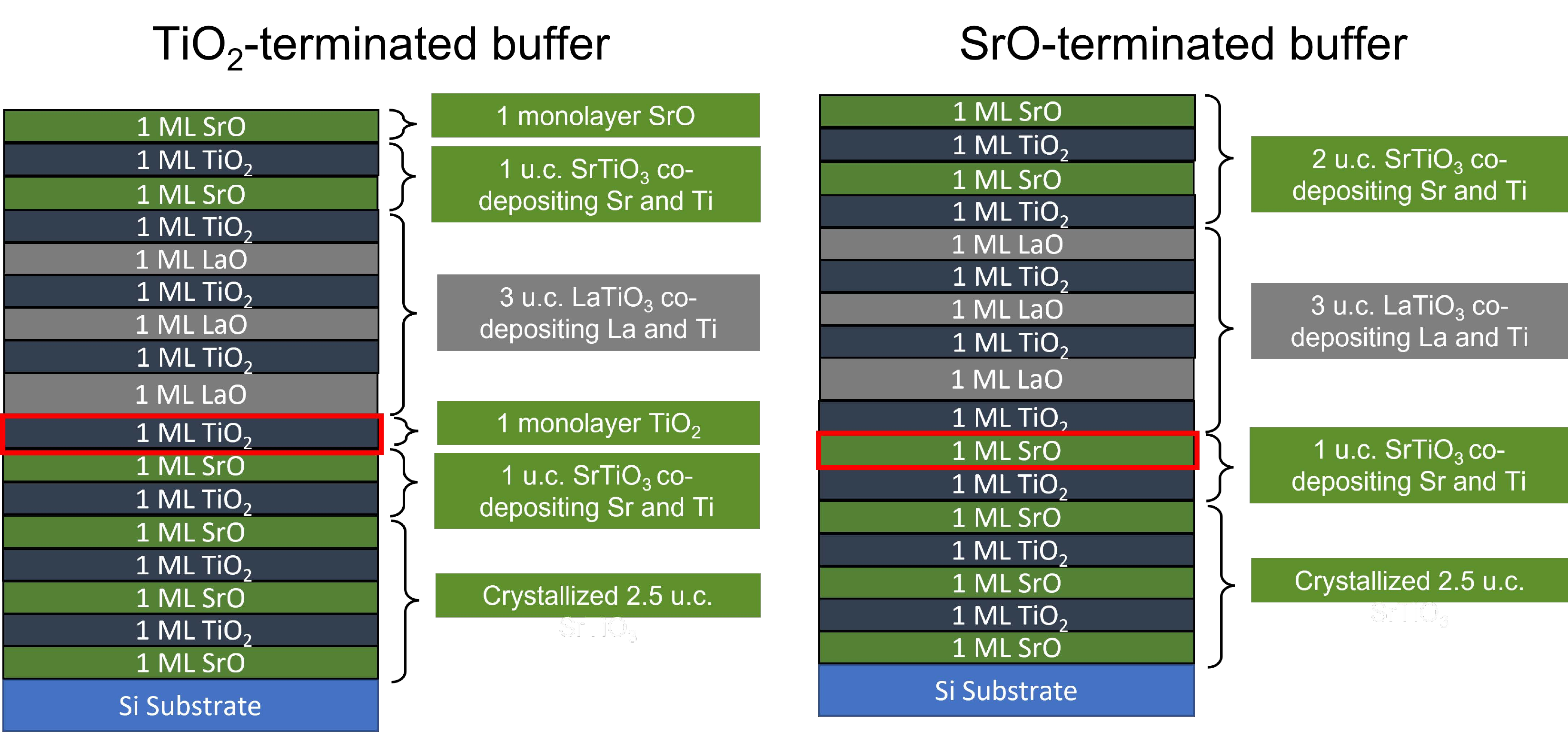}

  \caption{Schematics of 1.5 uc STO cap/ 3 uc LTO/ 4 uc STO/ Si samples grown by molecular beam epitaxy. Though identical in composition, the sequence of deposition and the terminating layer of the STO buffer (highlighted in red) prior to LTO deposition differ.}  
 \label{fig:i4schematic}
 
\end{figure}

Following the appearance of a 2 $\times$ 1 reconstruction in the reflection high energy electron diffraction (RHEED) pattern, indicative of a clean reconstructed Si surface, half a monolayer of Sr was deposited at 660 $^\circ$C to form a template for subsequent layers of STO. The substrate was then cooled to room temperature, and 3 ML of SrO and 2 ML of TiO$_2$ were co-deposited at room temperature and then heated to 500 $^\circ$C to form 2.5 ucs (two complete STO ucs terminated by an SrO layer) of crystalline STO, as shown in Figure \ref{fig:i4schematic}. Subsequent layers of STO and LTO of various thicknesses were grown at a substrate temperature of 500 $^\circ$C.

Two sets of samples were then grown for each STO buffer thickness as shown in the schematic in Figure \ref{fig:i4schematic} for the N=4 heterostructure. While identical in composition, the two sets differ in the sequence of deposition, and thereby the terminating layer of the STO buffer prior to LTO deposition. For the so-called TiO$_2$-terminated buffer, a single uc of STO was deposited on the 2.5 uc base layer of crystallized STO through co-deposition of Sr and Ti, followed by 1 monolayer of TiO$_2$ to form the buffer, as indicated in Figure \ref{fig:i4schematic}. Three ucs of LTO followed by 1 uc of STO were then deposited by co-deposition of La/Sr and Ti. A single monolayer of SrO was then deposited to cap the heterostructure. In contrast, for the so-called SrO-terminated buffer, 1 uc of STO, 3 ucs of LTO and 2 ucs of STO were deposited sequentially all through co-deposition of Sr/La and Ti fluxes on top of the 2.5 uc crystallized base, as indicated in Figure \ref{fig:i4schematic}. For both methods, the STO buffer was briefly annealed in vacuum at 580 $^\circ$C immediately prior to deposition of LTO, to enhance crystallinity. For the N = 6 heterostructures, samples with TiO$_2$-terminated and SrO-terminated buffers were deposited in the manner outlined above, except that the buffer has 2 additional ucs of STO in comparison to the N = 4 samples.

The transport properties of the trilayers were measured in the Van-der-Pauw configuration with Au contacts sputtered on the sample corners. The temperature-dependent sheet resistance was measured using a Keithley 2400 sourcemeter and a Keithley 2700 multiplexer in a Quantum Design Physical Property Measurements System (PPMS). 

To determine the relationship between the structural and transport properties of the films, the atomic-scale structure of the samples were determined by high-resolution synchrotron X-ray diffraction crystal truncation rod (CTR) measurements.\cite{robinson_crystal_1986, kumah_atomic_2010} X-ray diffraction measurements were performed at the 33ID beamline at the Advanced Photon Source.  The samples were mounted in a Be-dome chamber evacuated to a base pressure of 5$\times 10^{-5}$ Torr. The incident photon energy was fixed at 16 keV ($\lambda = 0.7749 \AA{}$). The diffracted intensities were measured using a Pilatus 100K 2D detector.\cite{schleputz2005improved}

\section{Results and Discussion}

\begin{figure}[ht]
\includegraphics[width=0.75\textwidth]{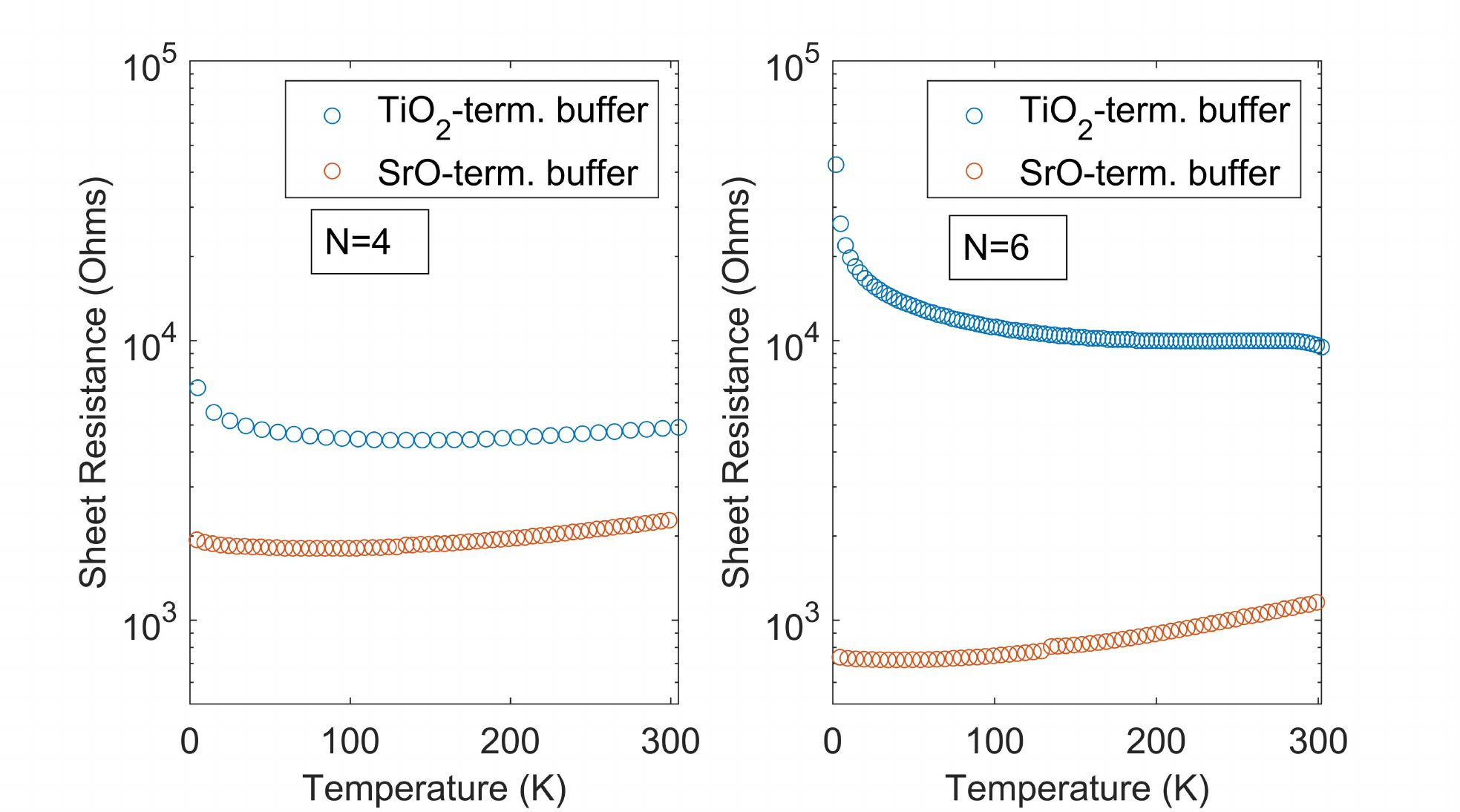} 
\caption{Comparison of the sheet resistance as a function of temperature for nominal 1.5 uc STO cap/ 3 uc LTO/ N=4,6 uc STO buffer/ (001) Si substrate heterostructures with SrO and TiO$_2$ terminated STO buffers.  }

\label{fig:I4transport}
\end{figure}

The temperature-dependent transport properties of the samples as a function of the STO buffer thickness (4 uc and 6 uc) and termination (SrO or TiO$_2$) are given in Figure \ref{fig:I4transport}. The N=4 samples show metallic behavior for both buffer terminations, however, the N=4 sample with an SrO-terminated buffer where La/Sr intermixing is enhanced  as discussed below, exhibits a lower sheet resistance compared with the sample with a TiO$_2$-terminated buffer. The sheet resistance of the N=6 sample with the TiO$_2$-terminated buffer increases as the temperature decreases.\cite{ahmadi-majlan_tuning_2018} For the corresponding N=6 sample grown on an SrO-terminated buffer, the resistivity at room-temperature is reduced by 1 order of magnitude at room temperature and decreases as the temperature decreases indicative of a metallic ground state. The reduced resistivity for the SrO-terminated buffer is postulated to be due to enhanced La-Sr intermixing leading to the formation of metallic La$_x$Sr$_{1-x}$TiO$_3$.\cite{tokura_filling_1993}

Crystal truncation rods along the Si substrate-defined reciprocal lattice vectors (1 Si reciprocal lattice unit (r.l.u.) = 1/5.43 \AA{}$^{-1}) $were measured to determine atomic structure of the coherently strained fractions of the oxide heterostructures. The diffraction data of the coherently strained fraction of the film was analyzed using the genetic-algorithm based GenX X-ray fitting program.\cite{bjorck_genx_2007} In addition to the CTRs, relaxed film peaks were observed at non-integer in-plane Si reciprocal lattice vectors. The in-plane lattice vectors of the incoherent fraction of the films do not coincide with the lattice vectors of the Si substrate, thus, reciprocal-space mapping measurements were performed to determine the lattice parameters of the strain-relaxed portions of the films.

\begin{figure}[ht]
\includegraphics[width=0.75\textwidth]{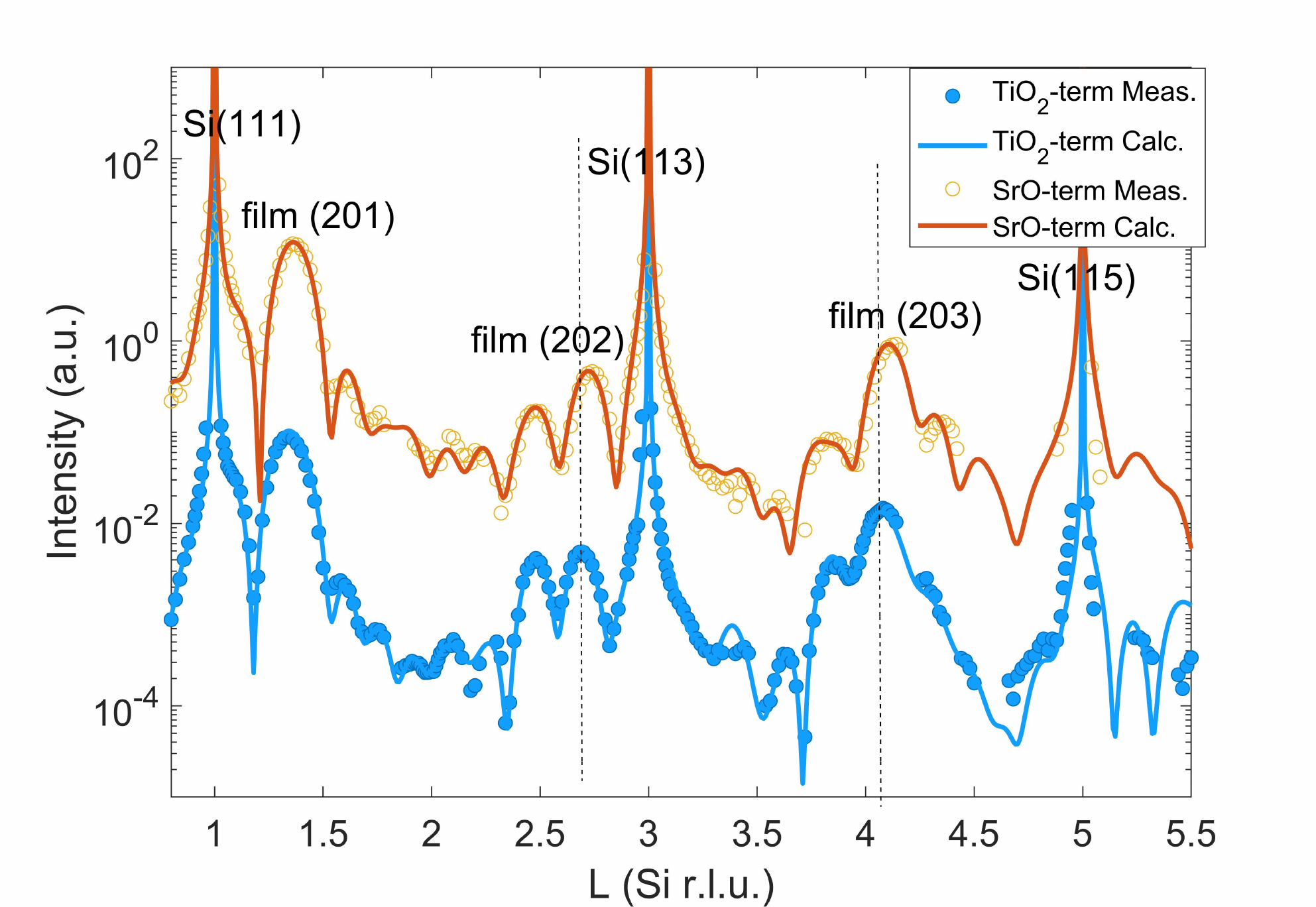} 
\caption{Comparison of the measured and fitted diffraction intensities along the Si 11L crystal truncation rods for nominal 1.5 uc STO cap/ 3 uc LTO/ 4 uc STO buffer/ (001) Si substrate heterostructures with SrO and TiO$_2$ terminated STO buffers. The dashed vertical lines indicate the locations of the Bragg peaks for the TiO$_2$-terminated buffer sample. }

\label{fig:I4compare}
\end{figure}

Figure \ref{fig:I4compare} shows a comparison of the measured crystal truncation rods along the Si H=1 K=1 direction for the N=4 heterostructures with SrO and TiO$_2$ terminated buffer layers. Due to the rotation of the perovskite unit cell by 45$^o$ relative to the Si lattice, the perovskite (20L$_{film}$) peaks are present along the Si (11L) crystal truncation rod.

A significant difference observed in the measured data for the 2 buffer terminations is the position of the film Bragg peak. The film Bragg peaks of the SrO-terminated sample are shifted to higher L values as compared to the TiO$_2$-terminated sample. This indicates a smaller (larger) average out-of-plane lattice spacing for the SrO (TiO$_2$) terminated buffer samples. The calculated average film lattice parameters for the SrO and TiO$_2$ terminated buffer samples are 3.95 \AA{} and 4.02 \AA{}, respectively. 

To determine the layer-resolved structural profile of the fractions of the sample coherently strained (i.e. in-plane lattice constant is the same as the substrate) to the Si substrate, the measured data were fit using the GenX X-ray fitting program.\cite{bjorck_genx_2007} The fit parameters are the lattice parameters of each layer and the La and Sr occupations of the A-site of the perovskite unit cell to account for La-Sr intermixing across the STO/LTO interfaces. The simulated CTR's for the best fit structures are shown as solid lines in Figure \ref{fig:I4compare}.

\begin{figure}[ht]
\includegraphics[width=0.5\textwidth]{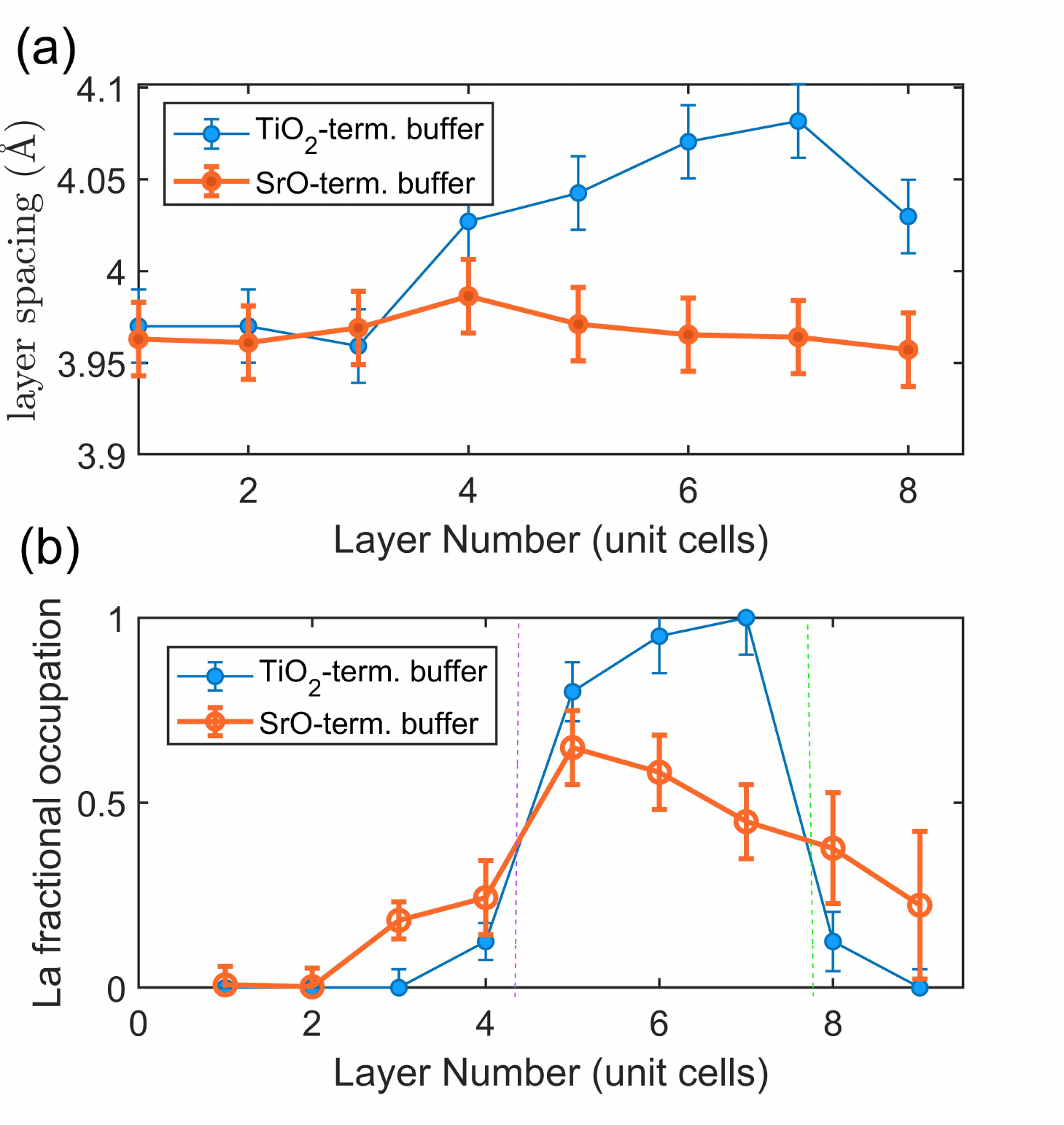} 
\caption{Comparison of the layer-resolved (a)lattice spacings and (b) La occupation profile for nominal 1.5 uc STO cap/ 3 uc LTO/ 4 uc STO buffer/ (001) Si substrate heterostructures with SrO and TiO$_2$ terminated STO buffers. The purple and green dashed vertical lines indicate the locations of the nominal STO buffer/ LTO and LTO/STO cap interfaces, respectively. }

\label{fig:I4structure}
\end{figure}

Figure \ref{fig:I4structure} shows a comparison of the layer-resolved out-of-plane lattice spacings and La/Sr chemical profiles for the 2 N=4 samples obtained from the CTR analysis. For both samples, the lattice spacings of the STO buffer layers (layers 1-4) adjacent to the Si substrate  are measured to be 3.96 \AA{} $\pm 0.01$ \AA{} corresponding to a c/a ratio of 1.03 (a=3.84 \AA{}). The layer spacings in the nominal LTO layers (layers 5-7) for the TiO$_2$-terminated (SrO-terminated) buffer sample have an average value of 4.05 $\pm 0.02$  (3.97 $\pm 0.02$) \AA{} corresponding to a c/a of 1.06 (1.034).

The composition profiles along the growth direction for the 2 samples are shown in the lower panel of Figure \ref{fig:I4structure}. While the La fractional occupation of the LTO layers in the TiO$_2$-terminated buffer sample are close to the expected value of 1, a significant reduction in the La content and a corresponding increase in the Sr content is observed within the LTO layers for the SrO-terminated buffer sample. The Sr-incorporation into the nominal LTO layer leads to the formation of metallic La$_x$Sr$_{1-x}$TiO$_3$ where the lattice volume decreases with the Sr concentration \cite{hays_electronic_1999} effectively reducing the lattice mismatch with the buffer layer. Thus, the reduced lattice constant observed for the SrO-terminated buffer sample is consistent with the measured chemical profiles.

\begin{figure}[ht]
\includegraphics[width=0.75\textwidth]{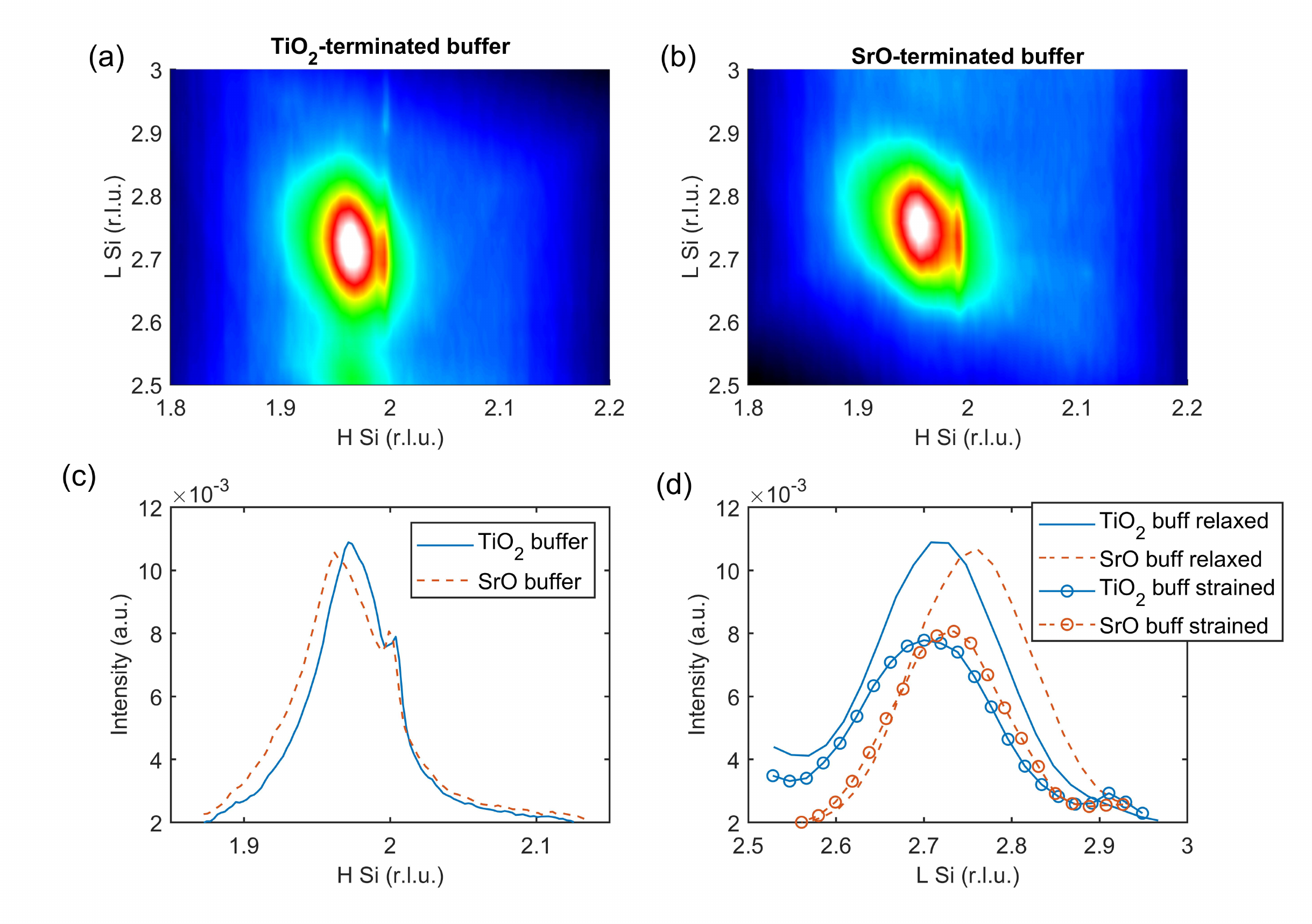} 
\caption{Reciprocal space maps around the (1 1 2) film peak for nominal 1.5 uc STO cap/ 3 uc LTO/ 4 uc STO buffer/ (001) Si substrate heterostructures with (a) TiO$_2$ and (b) SrO terminated STO buffers. (c) Line profiles in the in-plane H-direction through the maps in (a) and (b) for L=2.72 Si r.l.u. (TiO$_2$ termination) and L=2.76 Si r.l.u. (SrO termination). (d) Line profiles for fixed H in the L-direction through the strained and relaxed film peaks.}

\label{fig:I4RSM}
\end{figure}

The reciprocal space maps (RSM) around the H=2 K=0 L=2.7 Si r.l.u. are compared for the TiO$_2$ and SrO buffer terminated  N=4 samples in Figure \ref{fig:I4RSM}(a) and (b). Due to the epitaxial relationship between the Si and the perovskite film unit cell, the (2 0 2.7) Si peak corresponds to the (1 1 2) perovskite film peak. For each sample, 2 peaks are observed. The narrow peak along H=2 Si r.l.u. corresponds to the fraction of the film coherently strained to the Si substrate with the in-plane lattice parameter values of 3.84 \AA{}. The broader peak at lower H values corresponds to relaxed fractions of the film with in-plane lattice parameters larger than 3.84 \AA{}. The observation  of relaxed film peaks indicates that lateral strain relaxation occurs independent of the STO buffer termination. The average lattice parameters of the relaxed and strained fractions of the film can be determined from the peak positions in Figure \ref{fig:I4RSM}(a) and (b). Table \ref{tab:I4peak} summarizes the average lattice parameters calculated from the RSMs for the two samples.

Figure \ref{fig:I4RSM}(c)  shows a cut in the H-direction along K=0 and L=2.72 Si r.l.u. (TiO$_2$ termination) and L=2.76 Si r.l.u. (SrO termination). The peaks for the relaxed fractions are located at H=1.973  Si r.l.u. and H=1.962 Si r.l.u. for the TiO$_2$ and SrO terminated buffers, respectively.  The line profiles along the L direction for the strained and relaxed fractions are shown in Figure \ref{fig:I4RSM}(d). Here, the peak for the TiO$_2$-terminated sample is shifted to a higher L value than the SrO sample indicating a smaller out-of-plane spacing for the TiO$_2$ sample. 

\begin{table}
\begin{center}
\begin{tabular}{ | m{10em} | m{7em}| m{7em} | } 
\hline
\hline
Parameter                 & SrO-term  &  TiO$_2$-term    \\ \hline
H$_{strained}$(Si r.l.u.)   & 2.00         &  2.00   \\
a$_{strained} (\AA{})$      & 3.84         &  3.84  \\
\hline
H$_{relaxed}$  (Si r.l.u.)  & 1.962         &  1.973   \\
a$_{relaxed} (\AA{})$      & 3.915        &  3.893     \\
\hline
L$_{strained}$ (Si r.l.u.)     & 2.7308    &  2.7085   \\
c$_{strained} (\AA{})$        & 3.977     &  4.010      \\
\hline
L$_{relaxed}$ (Si r.l.u.)     & 2.7564      &  2.7181   \\
c$_{relaxed}(\AA{})$        & 3.9406      & 3.9962     \\
\hline
V$_{strained}(\AA{}^3)$       & 58.643      &  59.13   \\
V$_{relaxed}(\AA{}^3)$        & 60.398      & 60.56     \\
\hline

\hline

\end{tabular}
\caption{Comparison of lattice parameters for the relaxed and strained Bragg peaks for the TiO$_2$ terminated buffer and the SrO-terminated buffer sample.}
\label{tab:I4peak}
\end{center}
\end{table}


 When LaO and TiO$_2$ are co-deposited on a SrO-terminated STO buffer, La species can directly react with SrO layer to form an alloy which reduces the lattice volume and minimizes the strain energy of the system. The suppressed La/Sr exchange observed for LTO deposited on TiO$_2$ terminated STO buffer layers suggest that the TiO$_2$ layer serves as an effective barrier layer for La-Sr interdiffusion between LTO and STO. For the samples with the TiO$_2$ terminated buffers, as the STO buffer thickness is increased from N=4 to N=6, the epitaxial strain in the STO relaxes leading to a reduced lattice mismatch between the STO and LTO adlayer. Thus, strain-driven intermixing is expected to be further suppressed in the N=6 sample leading to more chemically abrupt LTO/STO interfaces and insulating behavior observed for the N=6 sample in Figure \ref{fig:I4transport}.\cite{ahmadi-majlan_tuning_2018}

\section{Conclusion}
In conclusion, we have demonstrated how strain-driven chemical intermixing at the LTO/STO interface is strongly dependent on the chemical termination of the STO buffer. Enhanced La/Sr intermixing for LTO co-deposited on SrO-terminated STO strained to Si leads to enhanced conductivity due to the formation of metallic LSTO. We find that the La/Sr exchange is suppressed for TiO$_2$-terminated STO buffers leading to a significant increase in the sheet resistance for samples with nominally identical composition. These results highlight the critical importance of chemical interactions driven by epitaxial strain and the composition of the interface terminal layer on the physical properties of functional oxide materials. We demonstrate that deposition sequence and terminating layers can be exploited to promote or minimize cation intermixing in layered heterostructures integrated on Si(100).

\section{Acknowledgements}
This work was supported by the National Science Foundation (NSF) under Awards No. DMR-1751455. Synthesis and transport measurements  were supported by the National Science Foundation (NSF, Award No. DMR-1508530). The use of the Advanced Photon Source was supported by the U. S. Department of Energy, Office of Science, Office of Basic Energy Sciences, under Contract No. DE-AC02-06CH11357.

The data that support the findings of this study are available from the corresponding author upon reasonable request.

%

\end{document}